\documentclass[journal]{IEEEtran}

\usepackage[latin1]{inputenc}
\usepackage{graphicx, amssymb, amsmath, amsfonts, amsthm}
\usepackage[usenames,dvipsnames]{color}

\allowdisplaybreaks
\IEEEoverridecommandlockouts

\begin{document}
\title{Wireless Information and Power Transfer:\\ A Bottom-Up Multi-Layer Design Framework}

\author{Ioannis Krikidis, \IEEEmembership{Fellow, IEEE}, and Constantinos Psomas, \IEEEmembership{Senior Member, IEEE} 
\thanks{I. Krikidis and C. Psomas are with the Department of Electrical and Computer Engineering, Faculty of Engineering, University of Cyprus, Nicosia 1678 (E-mail: {\sf \{krikidis, psomas\}@ucy.ac.cy}).}}

\maketitle

\begin{abstract}
The efficiency of wireless information and power transfer (WIPT) systems requires an essential reevaluation and rethinking of the entire transceiver chain, which is characterized by a bottom-up multi-layer design approach. In this paper, we introduce and describe the key design layers: i) ``Mathematical modeling'', associated with the investigation of mathematical models for the wireless power transfer process, ii) ``Information-theoretic limits'', which refers to the fundamental limits of the WIPT channel, iii) ``Link design'', corresponding to signal processing techniques that make WIPT feasible, iv) ``System-level perspective'', which studies the developed WIPT techniques from a macroscopic system-level point-of-view, and v) ``Experimental studies'', that refers to real-world implementation of WIPT systems. These layers are well-connected and their interplay is imperative for the effective design of WIPT systems. Specific case studies are discussed, which demonstrates the interdisciplinary nature of the aforementioned multi-layer design framework.
\end{abstract}


\section{Introduction}

Wireless information and power transfer (WIPT) is a new communication paradigm, which exploits the dual use of radio-frequency (RF) signals to convey information but also to energize low-power devices \cite{BRU1}. It appears as an attractive technology for the upcoming 6G communication systems, which are characterized by the massive connectivity of heterogeneous low-power devices, under the umbrella of the Internet of Everything and the Internet of Intelligence. The energy sustainability of these devices becomes a critical issue and so conventional solutions such as regular/frequent battery replacement and/or energy harvesting from natural resources are unpredictable, unstable, costly and in some scenarios infeasible. For this reason, through the co-design of information and energy signals, far-field WIPT can be a powerful alternative, which can ensure a continuous and fully-controlled energy harvesting process while at the same time enable communication connectivity. From the seminal work of Varshney \cite{VAR}, which introduced the fundamental concept of WIPT, a plethora of works have appeared in the literature, studying WIPT from different perspectives \cite{BRU1}. By categorizing the existing work, we can observe that WIPT is a multi-layer concept with specific questions and tools per layer. However, a systematic design approach that demonstrates the multi-layer nature of WIPT is missing from the literature.

In this paper, we present a bottom-up multi-layer framework for the design of WIPT systems. The considered approach takes into account that the WIPT channel comprises a linear information transfer channel and a nonlinear power transfer channel, along with the fact that the rectifying antenna (rectenna) circuit is an essential component for the end-to-end design. The bottom layer (Layer I) is the ``mathematical modeling'', which focuses on obtaining accurate and tractable mathematical models for the rectenna circuit. This layer serves as a basis for the fundamental study and development of WIPT. However, current efforts mainly simplify the characteristic function of the rectification process to linear/nonlinear wireless power transfer (WPT) functions. Ergo, the mathematical representation of the associated nonidealities is a key open challenge. To unlock the potential gains of the WIPT channel and study the fundamental trade-off between information and energy transfer (if it exists), the second layer relates to the WIPT ``information-theoretic limits'' (Layer II). By adopting the mathematical models from Layer I, this layer determines the limits of the WIPT channel, in terms of the information-energy capacity region as well as the associated input distributions and achievability techniques. Nevertheless, this layer does not provide any (practical) feasible techniques.

As such, Layer III deals with the ``link design'', where the signal processing and radio resource management design aspects of WIPT are examined. This third layer translates the information theoretic studies to feasible WIPT techniques and investigates waveforms, precoding, modulation schemes, receiver architectures, resource allocation methods etc. for various WIPT topologies. The limitation of this layer is that it mainly assumes deterministic and simple topologies, which are not able to capture the densification and spatial randomness that characterize modern wireless networks. To fill this gap, we take one step further-up to Layer IV, which adopts a ``system-level perspective'' and studies the integration of the developed WIPT techniques in complex networks (e.g., cellular, ad-hoc, etc.) with randomness on the spatial locations of base stations, users, obstacles, etc. With system-level studies, we can fully characterize the dual impact of multi-user interference, which on the one hand deteriorates the information performance, on the other hand enhances energy harvesting capabilities. The last and top layer ``experimental studies'' (Layer V), provides a proof-of-concept for the developed WIPT schemes and demonstrates their potential practical interest. By combining a communication platform with appropriate WPT circuits, this layer evaluates the accuracy of the theoretical models and the efficiency of the proposed WIPT techniques so that real-world feedback is provided for further improvement. Even though this layer is at the top of our design framework's hierarchy, it has a vertical interplay with all the previous layers.

\begin{figure*}\centering
	\includegraphics[width=0.8\linewidth]{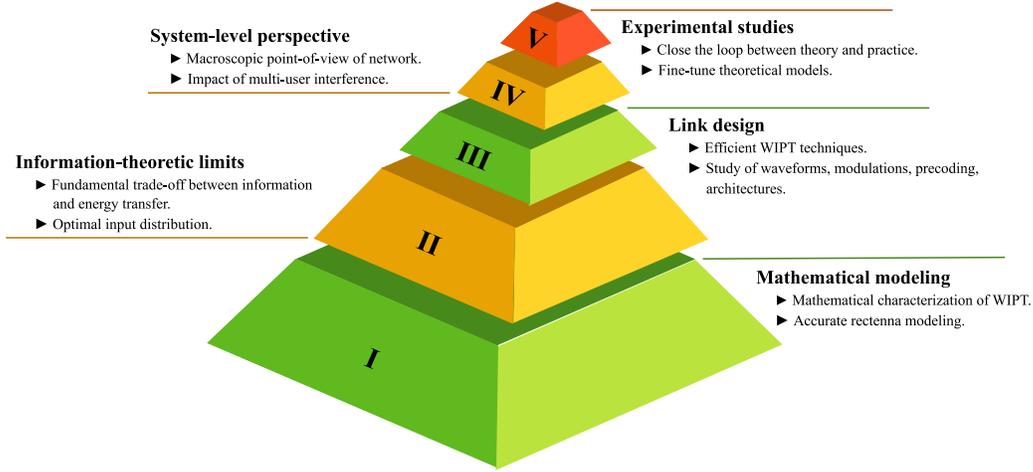}
	\caption{Bottom-up multi-layer design framework for WIPT.}\label{framework}
\end{figure*}

Fig. \ref{framework} schematically presents and outlines the bottom-up multi-layer WIPT design framework and a detailed description of each layer is provided in the next section.

\section{A Bottom-up multi-layer Design Approach\\ for WIPT Systems}

In this section, we introduce and describe the five design layers of the WIPT framework and discuss specific case studies. By following this bottom-up multi-layer approach, we achieve a close-to-reality WIPT design as well as a complete understanding of its potentials. 

\subsection*{\centering\rm\bfseries Layer I: Mathematical modeling}

The first layer for the design of WIPT systems refers to the investigation of mathematical models that capture the conversion of the RF signal to direct-current (DC) power. The key challenge is to propose mathematical models which are simple, tractable and at the same time highly accurate, as they constitute the first building block for the design of WIPT.

Inspired by the characteristic function of the rectification circuit (DC power versus RF power) and its three operation regimes (nonlinear, linear and saturation), three types of mathematical models dominate in the literature \cite{MOR}. The first one is the linear model where the DC power is a linear function of the received power and obviously refers to the linear regime of the characteristic function. A simple extension of this model is the piece-wise linear where an interconnection of linear segments approximates the behavior of the rectenna in a linear way. Although the linear models have been extensively used in the communication theory literature (mainly in the earlier research works in WIPT), they are not able to capture the nonlinear nature of the rectenna circuit and therefore result in inaccurate results. Another family of models tries to approximate the rectenna's characteristic function with a parametric nonlinear function (e.g., sigmoid, fractional, polynomial, etc.). By exploiting the extra degrees of freedom of these functions and by using curve-fitting tools, these models approximate more efficiently the WPT process. However, a key limitation of these models is the fact that their parameter adjustment refers to specific excitation signals and thus cannot be used to design waveforms and associated signal processing techniques. To overcome this limitation, circuit models have been proposed that apply fundamental circuit analysis to express the DC power as a function of the received RF signal \cite{BRU2}. These models are more appropriate for the design/optimization of WIPT systems but also suffer from several limitations i.e., they neglect several real-world phenomena such as impedance mismatching, parasitic effects, frequency/intermodulation product, RC filter, etc. The investigation of new mathematical models that overcome the current limitations is the foundation stone for the proper design of WIPT systems.

\begin{figure}\centering
  \includegraphics[width=\linewidth]{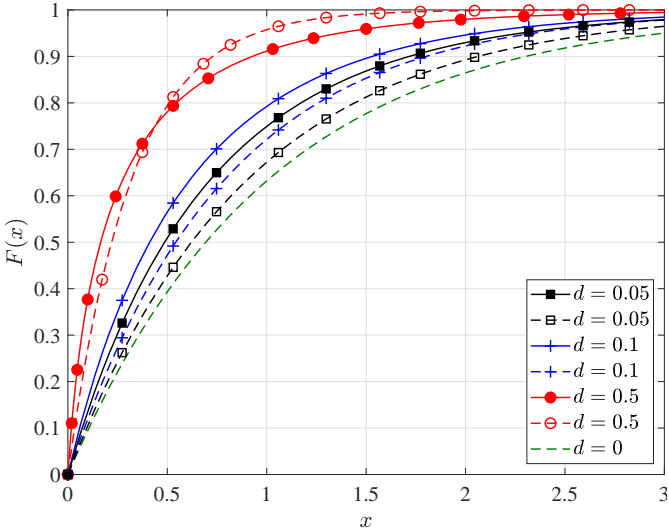}
  \caption{CDF of the actual distribution for different values of the divergence $d$; solid-lines correspond to the KL divergence between the distributions $f_0$ and $f$; dashed-lines correspond to the KL divergence between the distributions $f$ and $f_0$ (KL divergence asymmetry); the nominal distribution is normalized exponential.}\label{layer1}
\end{figure}

An efficient direction is to enhance the current deterministic mathematical models with some ``uncertainty'' that represents all these nonidealities, which are not taken into consideration; this uncertainty can be modeled by a partial knowledge of the WPT statistical distribution. By incorporating the concept of {\it compound channel}, we assume that the actual WPT statistical distribution $f$ is within a certain Kullback-Leibler (KL) divergence from a nominal distribution $f_0$; the nominal distribution $f_0$ and the maximum KL divergence $d$ determine the uncertainty \cite{KRI}. If the performance of the system is characterized by the average harvested energy, the associated (worst-case) optimization problem (i.e., minimize the average harvested energy with respect to the unknown distribution and subject to the KL constraint) is convex and can be solved with standard mathematical tools. As a toy example, in Fig. \ref{layer1}, we plot the cumulative distribution function (CDF) of the unknown WPT distribution for different values of $d$ for the case of a normalized exponential nominal distribution. The resulting CDF can be used to estimate the (worst-case) average harvested energy. By taking into account uncertainty, the results are more conservative in comparison to the nominal distribution (which represents the current models) but much closer to the real-world performance. 

\subsection*{\centering\rm\bfseries Layer II: Information-theoretic limits}

Once an accurate WPT mathematical model is established, the second layer studies the fundamental limits of WIPT i.e., the fundamental trade-off between information and energy transfer together with the associated input distribution \cite{MOR, AMO}. Specifically, by using tools from information theory, this layer studies the {\it Shannon information capacity under a minimum energy rate} by taking into account the energy harvesting function, given by Layer I. The main outcome is the information-energy capacity region, which is the closure set of all the achievable information/energy tuples. Recent research studies deal with this region for various WIPT channel models (e.g., additive white Gaussian noise (AWGN) channel, medium access control (MAC) channel, multiple-input multiple-output (MIMO) channel etc). Clearly, the characterization of the information-energy capacity region highly depends on the adopted WPT models. For instance, a classical point-to-point AWGN channel with linear WPT model does not present any information-energy trade-off (a symmetric complex Gaussian distribution maximizes both information/energy transfer), while the consideration of a circuit-based nonlinear model results in an information-energy trade-off and the optimal input distribution becomes zero-mean asymmetric complex Gaussian \cite{BRU1}.

\begin{figure}\centering
  \includegraphics[width=\linewidth]{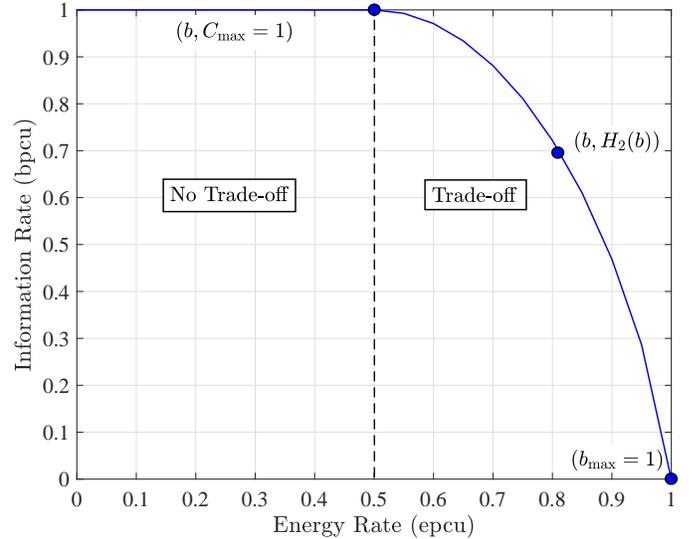}
  \caption{Information capacity under a minimum energy rate for a binary memoryless noiseless channel; the solid line shows the boundary of the information-energy capacity region.}\label{layer2}
\end{figure}

As a toy example, let us consider the (discrete) binary noiseless memoryless channel, which reproduces the binary input to the output without error. The information capacity of this channel is equal to $C=1$ bits/channel use (bpcu) and is achieved by a Bernoulli distribution with parameter $1/2$. Now, let us consider a WIPT framework where the binary symbol $1$ conveys $1$ energy-units/channel use (epcu) and the binary $0$ conveys $0$ epcu (e.g., similar to on-off keying modulation), while the minimum required energy rate at the channel output is $b$ epcu. If $b\leq 1/2$ epcu, the input distribution that maximizes the Shannon capacity (i.e., Bernoulli-$1/2$), also satisfies the energy rate constraint and hence there is not a trade-off between information and energy transfer. However, for the operation regime with $1/2 < b \leq 1$, the source transmits the binary symbol $1$ with a probability $b$ (Bernoulli-$b$) and therefore the associated information capacity becomes equal to $C = H_2(b)$, where $H_2(b)$ denotes the entropy of a binary random variable with Bernoulli-$b$; thus, for this operation regime, an information-energy trade-off is observed. It is worth noting that for the extreme cases where $b=b_{\text{max}}=1$ epcu, the Shannon capacity becomes zero ($C=0$) since the input random variable becomes deterministic. In Fig. \ref{layer2}, we plot the information-energy capacity region and the two operation regimes i.e., no trade-off regime ($b\leq 1/2$) and trade-off regime ($1/2<b\leq 1$). Other fundamental network structures such as the binary-symmetric channel, the erasure channel etc can be also considered with similar observations.

The fundamental limits serve as performance bounds to validate the efficiency of practical signal processing techniques (e.g., coding, waveforms, receiver architectures, etc.), which are designed at the third layer of the multi-layer design framework.

\subsection*{\centering\rm\bfseries Layer III: Link design}

This core layer includes the signal processing aspects of the WIPT design and summarizes the majority of the work in the literature. By using tools mainly from statistical signal processing and optimization theory, the key question here is to map the input distribution and the capacity achievable techniques (from Layer II) to physical signals/waveforms that approximate the theoretical bounds. Existing studies refer to different network structures and topologies and focus on specific signal processing aspects e.g., waveforms, modulation, precoding, receiver architectures, etc. \cite{BRU2, MUK, KIM}. 

\begin{figure}\centering
	\includegraphics[width=\linewidth]{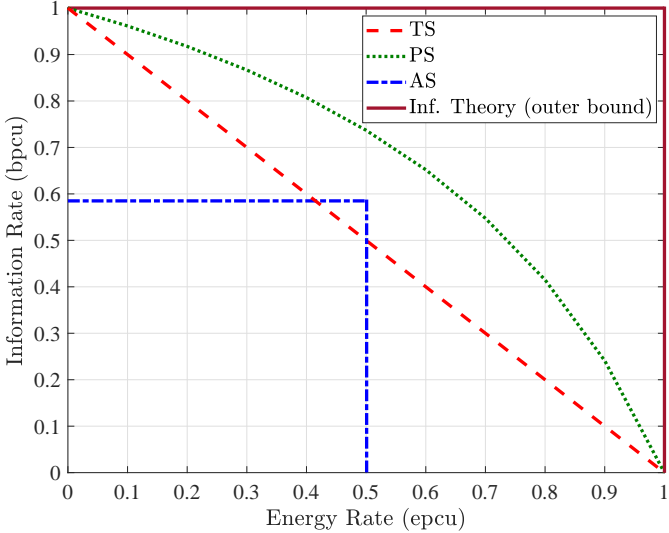}
	\caption{Information-energy capacity region for the conventional receiver arctitectures for WIPT; $1\times 2$ SIMO with $|h_1|^2=|h_2|^2=1/2$, $\sigma_n^2=\sigma_c^2=1/2$, PS with a symmetric splitting factor for each antenna.}\label{layer3}
\end{figure}

As a toy example to introduce Layer III and its interconnection with the previous layer, let us consider the three orthogonal WIPT receiver architectures: i) time-switching (TS), where the receiver switches in time between information and energy transfer, ii) power-splitting (PS), where the receiver splits the received signal into two components, one for communication and one for harvesting, and iii) antenna switching (AS), where the multiple antenna receiver partitions its antennas into two disjoint groups (one group of antennas is used for information transfer and the other for harvesting). In Fig. \ref{layer3}, we plot the achievable information-energy capacity region attained by these three techniques for a $1\times 2$ normalized single-input multiple-output (SIMO) deterministic channel with $|h_1|^2=|h_2|^2=1/2$ (channel power), transmit power $P=1$ W, AWGN with variance $\sigma_n^2=1/2$, RF-to-baseband noise with variance $\sigma_c^2=1/2$, and a linear WPT model; the information theory capacity bound is also presented. It can be seen that the PS technique outperforms TS (special case of PS with splitting factor $1/0$) and AS but all the practical schemes are far from the information theory (outer) bound due to the division of the resources between information and energy.

The challenge of closing this gap is still an open problem in the literature. A promising approach that recent studies take into consideration, is simultaneous information/energy transfer without (significantly) splitting the available resources. A key example is the {\it integrated receiver} \cite{ZHA}, where information is embedded in the amplitude of energy pulses and the conventional TS/PS/AS WIPT receiver is replaced by a simple rectification circuit without active RF chains. By taking samples at the output of the rectification circuit and through appropriate decision making mechanisms, the information is extracted accordingly; it is worth noting that an infinitesimally small power is allocated to the information receiver so practically almost all the received power is used for energy harvesting. Other techniques exploit the fact that signals with high peak-to-average power ratio (PAPR) such as multisine signals \cite{BRU2}, chaotic signals \cite{MUK}, etc. boost the energy harvesting performance (facilitate to overcome the building potential of the rectification circuits) and embed information bits in their physical characteristics. For instance, the works in \cite{KIM} and \cite{KRI2} embed information in the distinct values of PAPR and the number of sub-carriers, respectively, by using a pre-defined set of multisine signals.

It is worth noting that current works under Layer III focus on single blocks (e.g., coding or waveform design or precoding or reception schemes, etc.) and therefore a complete joint design is an open problem for Layer III. To overcome the difficulty of this model-driven approach, recent works adopt a data-driven approach and replace conventional transmitter/receiver chains with deep neural networks supported by data training (autoencoders) \cite{BRU1}.

\begin{figure}\centering
  \includegraphics[width=\linewidth]{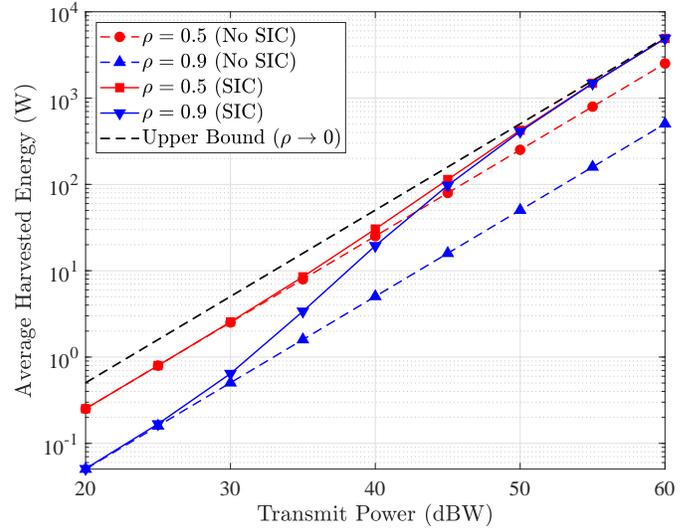}
  \caption{Average harvested energy in terms of the transmit power; PS parameter $\rho = 0.5$ or $\rho = 0.9$, density $\lambda = 10^{-3}$, and $d = 10$ m.}\label{layer4}
\end{figure}

\subsection*{\centering\rm\bfseries Layer IV: System-level perspective}

The design in Layer III concerns specific network structures with deterministic topologies and where randomness mainly concerns the fading channel statistics. Although these simple set-ups provide useful guidelines for the performance of real-world systems, they neglect the stochastic nature of the network topology, which characterizes modern dense network deployments. Hence, in Layer IV, we adopt a macroscopic point-of-view and study the designs of Layer III for large-scale networks that take into account the spatial randomness (associated with the location of the nodes), which is mainly embedded in the multi-user interference \cite{CP, CP2}. Specifically, by leveraging the mathematical models of random spatial point processes (e.g., Poisson, Ginibre, etc.) and tools from the stochastic geometry, the performance of the developed WIPT schemes can be studied for real-world dense networks (e.g., cellular, ad-hoc/sensor networks, etc.) by averaging out over all the potential network realizations. This mathematical approach allows the derivation of elegant closed-form mathematical expressions that characterize the performance of the entire network, by skipping the conventional complex and time-consumed system-level computer simulations.

The analysis of WIPT systems from a system-level perspective, reiterates and affirms the fundamental trade-off between information decoding and RF energy harvesting. In particular, high levels of multi-user interference degrades the performance of information transfer, while it becomes beneficial for energy transfer. Therefore, a reasonable direction is to find ways to exploit multi-user interference for energy transfer, whilst maintaining a desirable performance for information transfer. A potential approach is the employment of interference cancellation methods, such as successive interference cancellation (SIC), where a receiver attempts to decode the strongest interfering signals and, if successful, they are effectively removed thus increasing the signal-to-interference-plus-noise-ratio. In a WIPT context, SIC can be exploited to lower the PS parameter in such a way so that the achieved performance is still as good as the case where SIC is not applied. As the PS parameter decreases, more power is provided to the harvesting operation and so the average harvested energy is increased \cite{CP2}.

For example, consider a large-scale bipolar ad-hoc network consisting of a random number of transmitter-receiver pairs, where the transmitters
form a homogeneous Poisson point process (PPP) of density $\lambda$. Each transmitter has a unique receiver at a
distance $d$ in a random direction. Moreover, each receiver has simultaneous WIPT capabilities, using the PS method, and can perform SIC. Based on this scenario, Fig. \ref{layer4} shows the impact of SIC on the average harvested energy when the strongest interfering signal is canceled for different values of the PS parameter $\rho$. Clearly, the employment of SIC provides significant energy harvesting gains since an increase in the transmit power provides better quality signal, which results to more power for harvesting. For $\rho = 0.9$, the harvesting gains are noticeable from $25$ dB whereas for $\rho = 0.5$ from $40$ dB. This is again due to the fact that there is more than enough power at the receiver to achieve the desirable performance so the adjustment of $\rho$ starts from a lower transmit power. On the other hand, as the transmit power increases, the energy harvested with the proposed method converges to the upper bound ($\rho \to 0$).

\begin{figure}\centering
	\includegraphics[width=\linewidth]{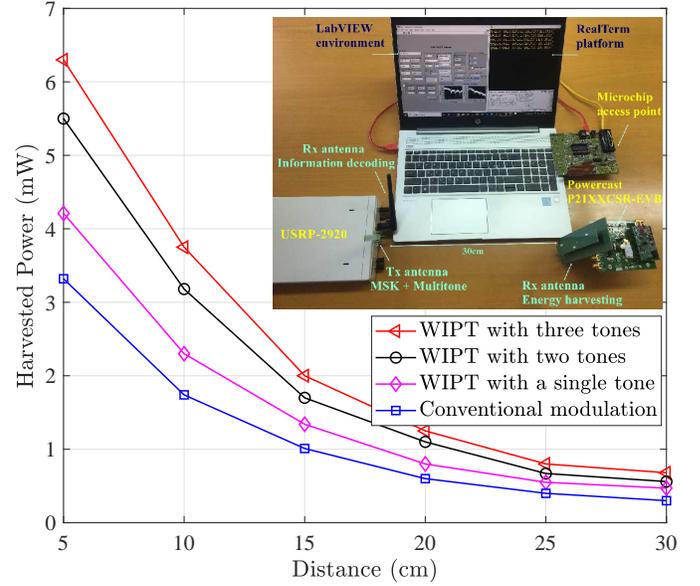}
	\caption{Experimental validation of the frequency-domain WIPT system by using an SDR-based communication platform and Powercast devices; transmission gain $30$ dB, RF carrier frequency $915$ MHz, sampling rate $1$ Msamples/sec; the information transfer performance is similar for all the schemes.}\label{layer5}
\end{figure}

\subsection*{\centering\rm\bfseries Layer V: Experimental studies}

An indispensable component of the WIPT technology is the rectification circuit and therefore the interplay between the theoretical models/techniques (Layers I-IV) and the hardware/implementation is vital. Layer V of the bottom-up design framework closes the loop between the theoretical and experimental studies and deals with the experimental validation of the developed techniques \cite{KIM2}. In addition, it acts as an umbrella layer as it provides real-world feedback to fine-tune and improve the mathematical models, the fundamental bounds as well as the developed communications techniques (i.e., close the loop between theory and practice and iterate). Although the implementation of rectenna circuits is a mature topic in the microwave/antenna literature, the development of a proof-of-concept for WIPT that bring theoretical ideas to real-world prototypes is still in its infancy. The key challenge is that the implementation of WIPT is a highly interdisciplinary task and requires a synergy of various areas of expertise such as wireless communication theory, microwave/antenna theory, RF engineering, etc.

Existing academic efforts mainly combine the flexibility of a software-defined radio (SDR) with customized or commercial rectennas (e.g., Powercast) in order to provide a low-cost information/energy hardware platform \cite{RAH}. For instance, the following experiment validates a frequency-domain WIPT waveform, where a multitone energy signal is embedded on the spectral nulls and/or low-power sub-bands of a conventional modulated signal (e.g., GMSK, QAM, PSK, etc.), in an indoor propagation environment. The hardware consists of i) a personal computer that displays the results of the experiment, ii) a universal SDR (USRP)-2920, which acts as the transmitter as well as the information decoding branch of the receiver, iii) a Powercast (P21XXCSR-EVB) evaluation board, which plays the role of the rectenna circuit, and iv) a microchip MRF24J40 access point, which communicates the measurements to the personal computer.

The inset image in Fig. \ref{layer5} shows the described experimental setup. In the experiment, we vary the distance between the transmitter and the receiver for a fixed USRP transmission gain at $30$ dB. Fig. \ref{layer5} depicts the harvested power as a function of transmitter-receiver distance for different waveforms corresponding to various numbers of multitones. It is observed that by superimposing more energy tones (for a given transmit power), we can harvest more power due to the high PAPR characteristics of the multitone energy signal, which improves the overall WPT conversion efficiency. By exploiting the fact that the energy tones are located in the spectral nulls (and/or low-power sub-bands) of the modulated information signal, the performance of the communication branch (e.g., bit-error rate performance) is not affected by the transmission of the energy signals. Therefore, the proposed WIPT waveform transfers simultaneously information and energy to the receiver, without splitting the available bandwidth and without any degradation in the communication quality.

\section{Discussion \& Open Challenges}

In the era of 6G networks where the need for connectivity and power sustainability further increases, WIPT can be considered as a building block to achieve ``zero-energy'' devices and provide extremely high energy efficiency. From the above discussion, it is clear that a narrow focus on a specific design layer, without taking into account its interaction with the other layers, results in suboptimal solutions with limited practical interest. The application of the presented systematic design framework demonstrates the multi-dimensional structure of the considered problems and seems to be a driving force to further advance WIPT technology. Specifically, by adopting this multi-layer design approach, we identify the following open challenges:
\begin{itemize}
\item Existing mathematical models are too simplistic and so are not able to capture all the physical phenomena, which emerge during the rectification and energy storage processes. Nonidealities related to hardware impairments as well as the impact of the operating frequency and/or the rectifier's topology (e.g., multistage Dickson's charge pump, DC/RF combiners) are not taken into account. The investigation of sophisticated mathematical models, which accurately represent the entire RF-to-DC process and provide a balance between the complex (but highly accurate) microwave theory models and the simplistic/tractable (but not very accurate) communication theory models is an open problem for Layer I.
\item Current works under Layer II are limited to basic network set-ups but also to simple WPT models and so the associated theoretical bounds have limited practical and design interest. The characterization of the information-energy capacity region for various network configurations (e.g., relay channel, broadcast channel, etc.) by taking into account advanced WPT models (proposed in Layer I), is the key challenge for Layer II. The consideration of more accurate WPT models is expected to provide different optimal input distributions and rate/capacity achieving schemes. The impact of channel state information (CSI) and/or feedback on the WIPT performance, the consideration of channels with memory as well as the fundamental limits for the finite blocklength regime are open problems for further investigation.
\item Equipped with accurate fundamental limits, the main challenge of Layer III is to convert the theoretical bounds to new feasible/practical WIPT techniques. The design of new waveforms which are beyond conventional shapes (e.g., multisine, chaotic signals) and exploit PAPR without deteriorating information transfer performance, is a key research direction for the feasibility of WIPT. In addition, advanced coding/modulation techniques as well as linear/nonlinear symbol-based precoding designs are important tools to achieve specific points of the information-energy rate/capacity regime. It is worth noting that the co-design of these three blocks through conventional optimization or machine learning tools in order to satisfy specific WIPT performance requirements (including RF exposure constraints \cite{ZHA}) is an interesting direction as well. On the other hand, the impact of CSI on the WIPT performance and the triple trade-off between channel estimation, information and power transfer as well as the interplay of WIPT with other enabling technologies (e.g., intelligent reflecting surfaces (IRSs), Terahertz communications, backscattering etc.) are also open problems for future work.
\item The advances made under Layer III need to be considered from a system-level standpoint and so Layer IV will attempt to answer the following questions: How can the developed WIPT techniques be modeled and analyzed when spatial randomness is taken into account? How these affect the distribution of multi-user interference and, as a result, the information/energy performance? Can these techniques be intelligently exploited from a system-level point-of-view (just like SIC in Fig. \ref{layer4})? Moreover, the emergence of 6G brings about new network topologies such as cell-free architectures, networks with unmanned aerial vehicles, IRS-aided topologies. Thus, the integration and suitability of WIPT with these types of networks should be explored and spatial point processes that efficiently capture their characteristics need to be considered. In addition, an effective unification of stochastic geometry with optimization theory and/or machine learning tools will be beneficial for the design of WIPT systems.
\item The experimental evaluation of WIPT techniques and architectures is still in its infancy and provides several opportunities for future research and development. Even though there are some prototypes that deal with (only) WPT or separate information and energy receivers, the implementation of co-located WIPT through conventional splitting (i.e., TS/PS/AS) and/or sophisticated waveforms and transmissions techniques (proposed in Layer III) is a key challenge. To have a significant progress in this layer, we need customized implementations (beyond the utilization of SDR and commercial rectennas), which require interdisciplinary research that combines expertise from antenna design, RF electronics, microwave and communication theory. A mature Layer V can demonstrate all the potentials of WIPT technology and facilitate its integration and commercialization in future 6G applications.
\end{itemize}

\end{document}